\documentclass[trackchanges]{aastex701}
\usepackage{float}

\newcommand\latex{La\TeX}
\begin{document}
\title{
Are carbon deflagration supernovae triggered by dark matter ?}

\author{Jeremy Mould}
\affiliation{Swinburne University}
\affiliation{ARC Centre of Excellence for Dark Matter Particle Physics}
\email{jmould@swin.edu.au}

\begin{abstract}

Collisions between stellar remnants and dark matter in the Galactic bulge are frequent, and the kinetic energy of a
primordial black hole incident on a white dwarf, if it is all thermalized, will raise the degenerate core's
temperature, by at least a degree in the case of a lunar mass black hole. This is an underestimate in two ways:
the specific heat is less than 3k/2 per particle, and the incoming object is accelerated by gravitational
focusing. Detailed physical models have recently been made of this triggering event. Present observational
data are equivocal as to whether the radial distribution of type Ia supernovae in galaxies follows the starlight
in the galaxies, or is more concentrated towards the center, as collisional triggering would suggest.
But future samples of millions of supernovae from the Rubin telescope will change that.
\end{abstract}

\newcommand{\kms}{\mbox{km\,s$^{-1}$}}
\newcommand{\etal}{\mbox{\rm{et al.}~~}}

%{\color{red}Draft 4 Feb 2026}

%\section*{Abstract}

\section{Introduction}
Carbon deflagration supernovae (SNe) are one of the best predictions of stellar astrophysics. They are also a pillar of 
observational cosmology.
But does the degenerate core of the Chandrasekhar mass white dwarf spontaneously explode, or is it triggered ?
This question is almost as old as electron degeneracy, but simulations of a triggering process have recently caught up (Leung \etal 2025).
Earlier work is by Graham \etal (2015). In this paper we examine the observational evidence that exists for dark matter (DM) triggering of SNeIa. 
If triggering by primordial black holes (PBHs) occurs, the best place to look for SNeIa would be the centers
of galaxies, where the peak density of DM is found.
We also investigate the number density squared rate dependence
of dark matter triggering in galaxies from the Millennium simulation.

\section{Observations}
The Dark Energy Survey (DES, Troxel \etal 2017) and OzDES\footnote{http://mso.anu.edu.au/ozdes} recorded 
and measured redshifts for 1829 type Ia SNe. Toy \etal (2023), using a redshift independent measure of radial distance
of the SN distance from the center of each galaxy, published the data in Figure 1.
This measure is also independent of surface brightness dimming (see Appendix).
\begin{figure}
	\includegraphics[width=\textwidth]{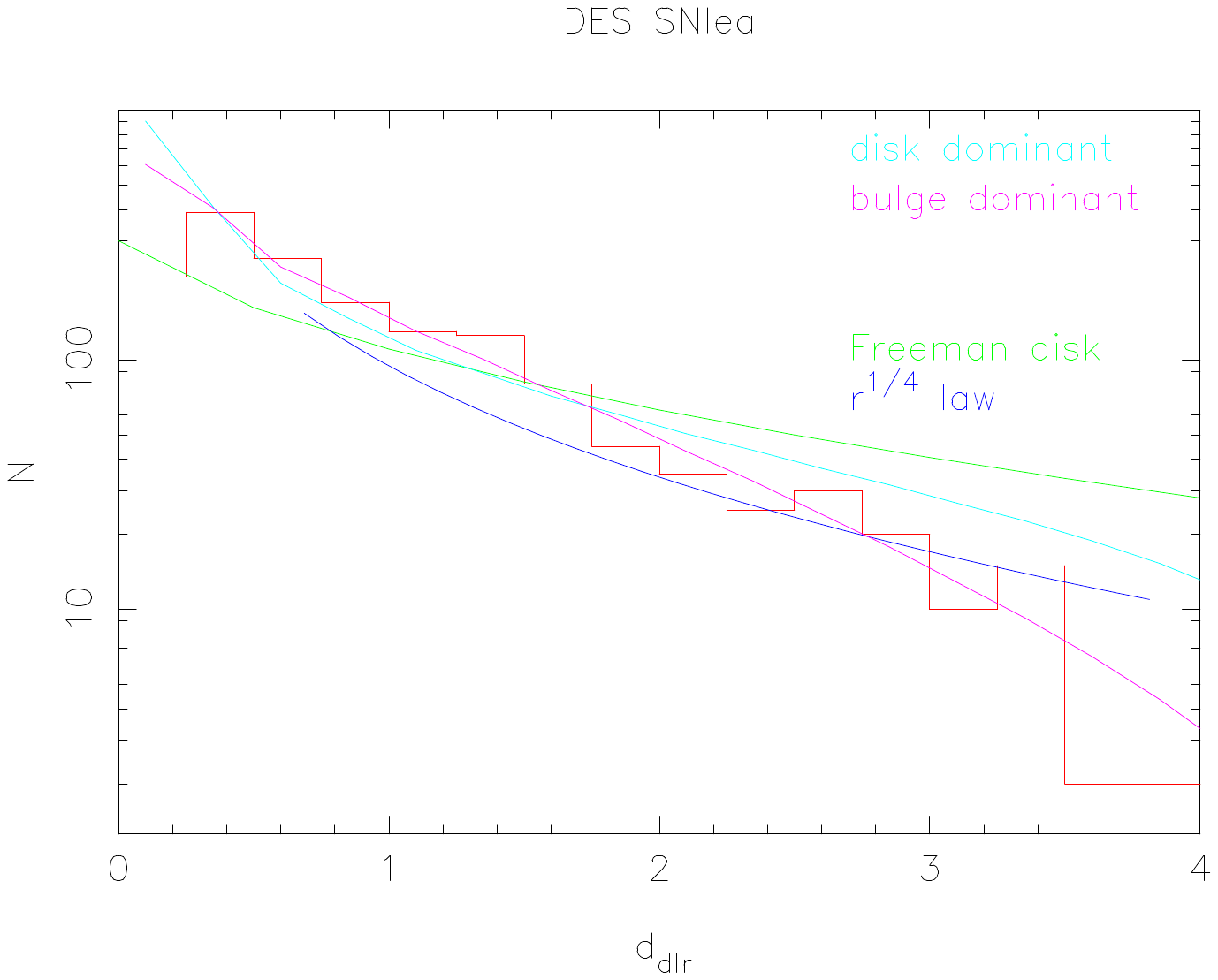}
	\caption{Radial distribution of SNeIa in the Dark Energy Survey (in red).
	The green curve is the expectation for exponential disks and the blue curve for elliptical galaxies.
	These have not been normalised to the data. The other two curves are for DM triggered SNe,
	and these have been normalized to the same total number of SNe, excluding the first, central bin,
	which is anomalous, and is conceivably a selection effect of some SN being masked by the nucleus.}
\end{figure}

To compare this with expectations from theory we need to relate the radial quantity d$_{DLR}$ with the exponential disk
(Freeman 1970) of spiral galaxies and the r$^{1/4}$ law of ellipticals (de Vaucouleurs 1959). The quantity d$_{DLR}$
is the second moment of the radial distribution, and for exponential disks this is 2a$^2$ - a for disk scale radius a.
Therefore 4d$_{DLR}$ = 1 + $\surd$ (8a). If the radial SN distribution follows the light of the old stellar population,
then the green curve is the expectation. For the r$^{1/4}$ law, numerical integration is required to obtain the second moment,
and the result is the blue curve in Figure 1.

In a galaxy with white dwarf number density n$_*$ and dark matter number density n$_\bullet$, the rate per unit volume of collisions is n$_*$ n$_\bullet~\sigma$ v,
where $\sigma$ is the collision cross section and v is the relative velocity. For spontaneous deflagration the same rate is just n$_*$. %A summary of the contents of the bulge of the Milky Way is given in Table 1.
%The prediction for triggered events is a higher central concentration of SNeIa than core collapse SNe, type II. Figure 1 shows that this prediction
%is borne out in the Harvard CBAT database.
%\begin{figure}
%	\includegraphics[width=\textwidth]{pgplot.pdf}
%	\caption{The radius distribution of SNeIa and SNeII in the CBAT database 1900-2015. The SNeIa distribution is in black and the SNII distribution
%	is in red. The numbers of these SNe are visible to the right.}
%\end{figure}
%The radii were calculated by multiplying distance estimates from the recorded SN magnitude plus 19 by the SN offset in arcsec from the galaxy's centre.
\begin{table}[H]
	\caption{Galactic bulge contents}
	\begin{tabular}{lllll}
		\hline
		Species&BH      &NS  & WD & MS\\
	Initial	Mass & max, min & min& min & min\\
		M$_\odot$&100, 3.5&1.17&0.7&0.1\\
	$\int$ n(m) m dm& log(100/3.5)&log(3.5/1.17)&log(1.17/0.7)&log(7)\\
	Mass 10$^{10}$ M$_\odot$   &1.60/3 & 0.52/3 & 0.25/3& 0.93/3\\
$\int$ n(m) dm&3.5$^{-1}$-0.01&1.17$^{-1}$-3.5$^{-1}$&1.43-1.17$^{-1}$&10-0.7$^{-1}$\\
	N 10$^{10}$&0.28/3&0.57/3&0.58/3&8.57/3\\
		\hline
		\multicolumn{5}{l}{MS = main sequence, NS = neutron star}
	\end{tabular}
\end{table}

\section{Triggered supernova rates}
The Galactic bulge is the densest region of our Galaxy, and so the most favorable for high collision induced SN rates.
The make-up of the Galactic bulge, an old stellar population, is given in Table 1 for an initial mass function n(m) $\propto$ m$^{-2}$, close to the Salpeter function. The mass of the bulge 1.1 $\times$ 10$^{10}$ M$_\odot$ is taken
from Binney \& Vasiliev (2023).
The rate per unit volume of collisions between these objects is
$$ \rho =  n_1 n_2 \sigma v~~~~~~\eqno(1)$$
where $\sigma$ is the cross section, v is the velocity, V is the volume and n$_1$, n$_2$ are the number densities of the interacting objects. 
 Gravitational focussing increases the probability of collision by a factor\footnote{Na\"ively, this factor
 can be seen as the ratio of solid angles of the two velocity ellipsoids} of g = (
1 + v$_{esc}^2$/v$^2$) (Barnes 2011,  Seligman, Marceta, \& Pe\~na-Asensio 2026), where v$_{esc}^2$ = 2 G (M$_1$ + M$_2$)/(R$_1$ + R$_2$). The small denominator of g yields a value of g of 4.8 $\times$ 10$^5$. %The volume of the bulge
If PBHs are a fraction f of the Galaxy's dark matter and a density
of 0.01 to 0.1 M$_\odot$ pc$^{-3}$ can be assumed for the Galaxy's dark matter core
(Cole \& Binney 2016, Lazar \etal 2020), then for 
a lunar mass PBH (10$^{-7}$ M$_\odot$) equation~1 gives a rate of 12.6 to 126 fv = 13.8 to 138 f $\surd$3 $\times$ 10$^2$  Myr$^{-1}$ for a 1D DM velocity dispersion of 110 \kms.
The
density of L* galaxies (like the Milky Way) is approximately
(Blanton \etal 2003, Bell \etal 2003) 0.01 h$^3\times$ Mpc$^{-3}$ = 3.9 $\times$ 10$^6$ Gpc$^{-3}$ for $h$ = 0.73 (Riess \etal 2024).
%WDs Rate = 
%0.58 f g / 3 10$^{15}$ $\times$ 9.6 $\pi$  $\times$ 10$^{-26}$ $\times$ 110 per Myr = 6.4 $\times$ 10$^{-9}$ Myr$^{-1}$. %\\ 
%How much energy? KE = 10$^{26}$ (110 10$^5)^2$ =1.2110$^{40}$ ergs 
%GRB250702B 10$^{54}$ ergs 
%mc$^2$ energy for WD is 10$^{33}$ 9 10$^{20}$ ergs = 910$^{54}$ ergs 
Over a cosmic volume this rate becomes 0.9 to 9 f $\times$ 10$^{4}$ Gpc$^{-3}$ yr$^{-1}$, compared with the observed rate of (2.55 $\pm$ 0.12) $\times$ 10$^4$ yr$^{-1}$ Gpc$^{-3}$ h$^3_{70}$ 
(Desai \etal 2026).
For f $\sim$ 20\% of the dark matter this is consistent with the rate estimate from observation. This value is close to upper limits from microlensing observations. Two other parameters need mentioning. Leung \etal (2026) ran white dwarf models down to 0.7 M$_\odot$. Our triggered SN rate would reduce by an eighth if the 0.7--0.8 M$_\odot$ failed to explode or burn. And sub-lunar mass PBHs would produce more SNeIa, and that can be compensated by reducing the parameter f, the fraction of the DM in PBHs.

To calculate the radial distribution of SNeIa triggered by PBH collisions, one can assume the NFW profile for the DM
and use the following quantities provided for the Millennium simulation Springel \etal 2005) by the Theoretical Astrophysical Observatory
(TAO: Bernyk \etal 2016): Stellar mass, bulge mass, DM halo mass, disk scale radius, and virial radius.
The surface density of SNeIa is then the integral of the volume density along the line of sight through the galaxy.
$$n(r) = \int \rho(s) dr^\prime     \eqno(2)$$
where s = $\surd$(r$^2$ + r$^\prime 2$). The ratio of the virial radius to the scale radius in the NFW profile
was assumed to be 12.

The acceptance criterion for TAO galaxies was stellar mass  10$^9~<$ stellar mass $<~10^{10.5}$ and satellite galaxies were
rejected. Of these, every fifth galaxy was selected, and 8811 disk dominated galaxies were chosen (disk mass $>$ bulge mass) and 2146 bulge dominated galaxies.
The results were averaged and plotted as labeled in Figure 1. 
There is considerable variance between galaxies, but the statistics have shrunk the error bars to
about the width of the plotted curves. 
By eye, only the pure (untriggered SN) Freeman disk is a poor fit.
%\section{The black hole mass}
%The model fitted by Eyres-Ferris \etal to the event is parameterized by
%an energy release of
%$$E = 2\pi R_{WD} l_C \DeltaR \rho v_s^2 \sim 2\pi l_C (R_{WD}^3 v_p^3 c \rho/\kappa)^{1/2} ~~~~\eqno(3)$$, where $l_C$ $\sim$ R$_{WD}$/c is the light crossing time and the pericentre velocity of the orbit, v$_p$ = (G M$_{BH}$/r$_p$)$^{1/2}$. When r$_p$ is replaced by the tidal radius, v$_p~\propto$ M$_{BH}^{1/3}$
%and E $\propto~\surd$M$_{BH}$. The MCMC fit of the model to the x-ray data yields
%log M$_{BH}$ = 4.02 $\pm$ 0.63 M$_\odot$.
\section{Historical supernovae}
Supernovae since 1930 have been recorded by
the Central Bureau for Astronomical Telegrams\footnote{http://www.cbat.eps.harvard.edu} and it was possible\footnote{More recent SNe are available by subscription.}
to download those before 2015. The offset of the SN from the center of the galaxy is frequently recorded in arcsec.
These can be made distance independent by dividing by the D$_{25}$ diameter, available in many cases from the NED database.
For a Freeman disk the relationship is a = D$_{25}$/6.8 $\times$ 2.5/ln(10). The values of D$_{25}$ are subject to (1+z)$^4$ surface brightness dimming with redshift z. I correct for this by dividing the diameters by (3.4-10z)/3.4, the numerical value 3.4 being the difference between the central surface brightness of the disk, 21.6 B mags per arcsec$^2$, and the 25th mag isophote. For the historical SNe at low z this is a small correction, as shown in Figure 2 (upper left).
Figure 2 shows that such a disk is a good fit to the radial distribution, with no sign of the excess central concentration
seen in Figure~1.

The lower two panels of Figure 2 show galaxies with K band 2MASS isophotal diameters. Their correction for surface brightness dimming and their transformation to D$_{25}$ by dividing by 0.69 is indicated in the Appendix.
\begin{figure}
	\includegraphics[width=1.2\textwidth]{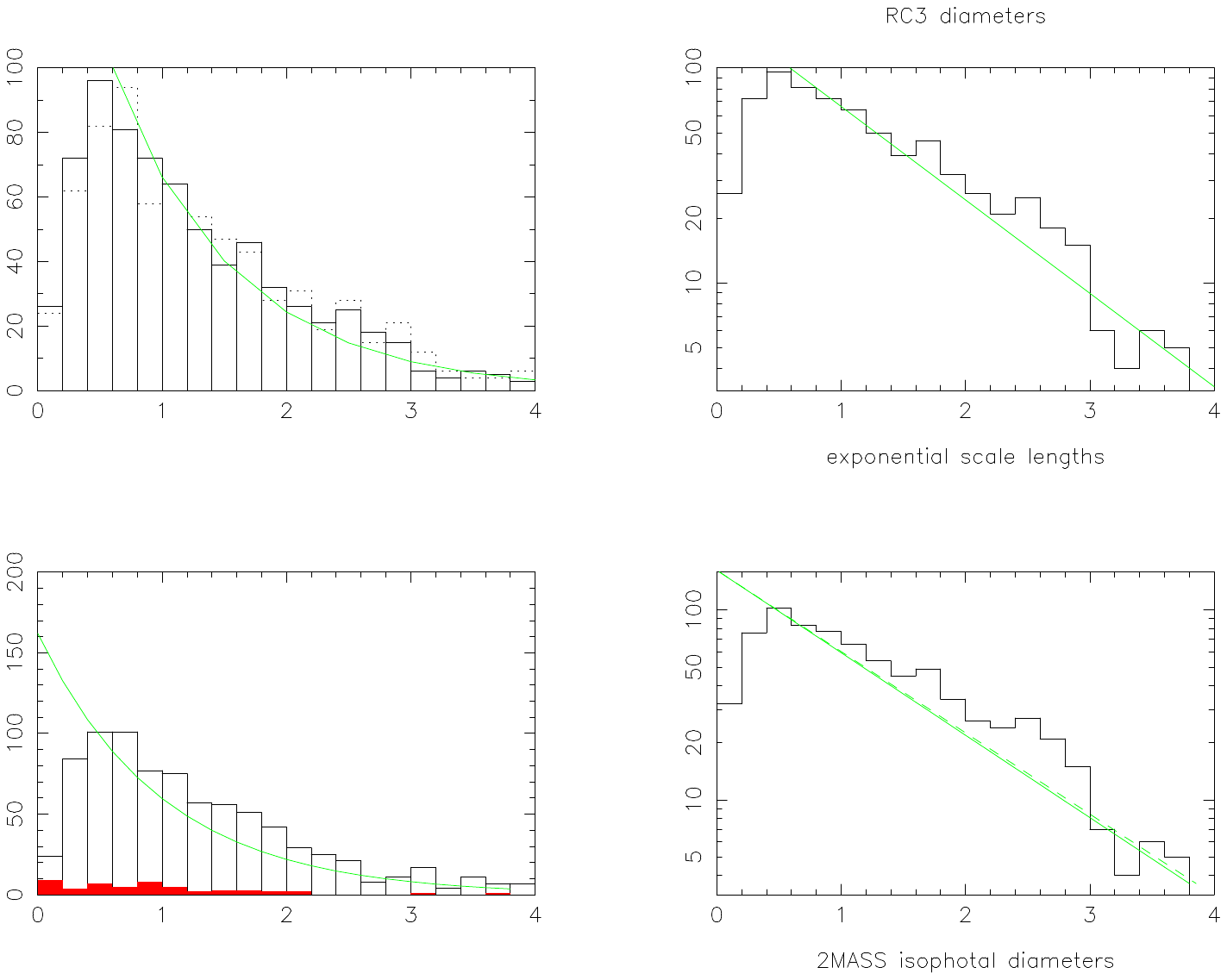}
	\caption{The top two panels are the radial distribution of SNeIa scaled to remove distance 
	dependence as described in the text, using the RC3 diameter D$_{25}$ (de Vaucouleurs \etal 1976). The solid histogram (upper left) has been corrected for (1+z)$^4$ surface brightness dimming, and the dotted histogram has not. The difference is small for these low z galaxies.
	The lower two panels use the total magnitude K band radius from the 2MASS survey.
	The shaded red histogram is the galaxies in Table 2 shown separately.
	In the bottom right hand plot they have been added in. In general the slope of these histograms is shallower than the expectation from the stellar light (the green curves and lines).}
\end{figure}
The historical SNe do not support the higher central concentration seen in the DES SNe.
No correction has been made for selection effects in either survey, and this is an area where
future work can improve confidence in the results. Table 2 contains host galaxies indicated as anonymous by CBAT
which were linked with their SNeIa by NED using a radius of 0.18$^\prime$. The column labeled 2MASS
lists their 2MASS isophotal diameters. 
\section{The Rubin Observatory}
This is an exceptionally well suited problem for Rubin telescope surveys to solve. Images of every recorded SN and the
surface brightness profile of its host galaxy provide all the data needed to repeat the work outlined here with
definitive effect. This may permit issues with galaxy surface brightness fitting to be avoided. Note that it is the surface brightness of the old stellar population that is relevant to these SNeIa, and that $\S$3 showed that the central density of the DM is the largest uncertainty in any rate prediction for triggered SNe.
Key data to be recorded are the surface brightness of the position of the SN in the post-SN galaxy, the SN type and the galaxy redshift.
Selection effects masking nuclear SNeIa need to be carefully modeled.
\section*{Acknowledgements}
My thanks to Xiuqin Wu for advice on the NED queries in $\S4$ and to Darren Croton for help with TAO. This research has made use of the NASA/IPAC Extragalactic Database (NED) which is operated by the California Institute of Technology, under contract with NASA  Award Number 80NSSC21M0037\footnote{10.26131/IRSA97} . The Central Bureau for Astronomical Telegrams and the Minor Planet Center are services of the International Astronomical Union and operated by Dept. of Earth and Planetary Sciences, Harvard University.
%Harvard SN catalog. 
The 2MASS project is a collaboration between The University of Massachusetts and the Infrared Processing and Analysis Center (JPL/ Caltech). Funding was provided primarily by NASA and the NSF. 
TAO is part of the All-Sky Virtual Observatory (ASVO) and is funded and supported by Astronomy Australia Limited, Swinburne University of Technology and the Australian Government. The latter is provided through the Commonwealth's Education Investment Fund and National Collaborative Research Infrastructure Strategy.
%\documentclass{article}
%\newcommand{\kms}{\mbox{km\,s$^{-1}$}}
%\begin{document}
\begin{table}
\vspace{-0.9in}
\caption{Deanonymized galaxies from CBAT}
\begin{tabular}{lrlrr}
\hline
SN         & r   &  Host                    & 2MASS    &  cz\\
           &arcsec&&                          arcsec   &\kms\\
 \hline
2015ax     & 12.2& WISEAJ011814.92+150556.7& 20.8      &      \\
2014ci     &  4.5& WISEAJ223404.03+682630.3& 78.6      &  18748    \\
2014ch     &  1.4& WISEAJ155830.83+125155.3& 15.4      &  43550    \\
2014am     &  6.1& WISEAJ152318.43+181728.1& 17.4      &  13724    \\
2013ew     & 32.0& WISEAJ221009.32+111657.1& 27.0      &  20066    \\
2012da     & 22.8&WISEAJ130233.83+272610.4& 22.2      &  20985    \\
2012ae     &  3.2& WISEAJ085856.06+230341.0& 21.0      &  14227    \\
2011hz     &  1.0& WISEAJ085234.32+551513.5& 11.4      &  10283    \\
2010lk     &  2.2& WISEAJ091514.39+012213.4& 52.4      &  98400    \\
2010ix     & 14.6&WISEAJ010002.74+414306.5& 25.6      &  67040    \\
2009mp     &  0.0&WISEAJ092359.89+140646.9& 23.0      &  33429    \\
2009jm     &  1.0& WISEAJ225014.07+103333.8& 20.4      &  16679    \\
2009hw     &  2.2& WISEAJ192033.98+433349.2& 13.6      &  86940    \\
2009ez     &  0.0& WISEAJ131254.43+432836.0& 13.8      &  23974    \\
2009eg     &  7.1& WISEAJ145415.48+185746.7& 13.6      &  13629    \\
2009df     &  0.0& WISEAJ160553.08+172438.8& 10.0      &  17456    \\
2009co     &  6.1& WISEAJ122435.85+471415.2& 12.2      &  65954    \\
2009cm     &  0.0& WISEAJ115442.17+551810.6& 10.0      &  48962    \\
2009aw     &  6.4& WISEAJ063146.75+245509.5& 13.0      &  42113    \\
2006eq     &  9.9& WISEAJ212837.50+011346.3& 16.2      &  77950    \\
2006da     &  4.5& WISEAJ232748.75+142831.1& 20.6      &  81050    \\
2006ct     &  3.2& WISEAJ120956.71+470545.5& 14.2      &  12324    \\
2006cj     &  4.5& WISEAJ125924.11+282049.8& 14.2      &  94280    \\
2006cg     &  3.6& WISEAJ130502.54+284420.4& 25.2      &  20298    \\
2005hf     &  2.8& WISEAJ012705.85+190701.8& 19.8      &  20327    \\
2005eu     &  1.4& WISEAJ022743.32+281037.7& 24.2      &  12924    \\
2005be     &  7.1& WISEAJ145933.08+164006.8& 17.6      &  10370    \\
2003hw     &  3.6& WISEAJ030149.87+354435.2& 22.6      &      \\
2003ay     &  7.3& WISEAJ040726.44+280748.4& 12.8      &  10053    \\
2003av     &  4.1& WISEAJ080132.37+024825.9& 11.6      &  16189    \\
2002eu     & 15.6& WISEAJ014943.12+323736.3& 124.8      &  11627    \\
2002lq     &  4.1& WISEAJ164028.34+411409.7& 26.0      &      \\
2001bp     &  7.2& WISEAJ160208.91+364313.9& 14.2      &  39880    \\
1999ax     &  3.6& WISEAJ140357.57+155111.7& 21.2      &  10116    \\
1998aa     & 10.0& WISEAJ122532.63+073859.6& 21.8      &  35836    \\
1997ea     &  9.5& WISEAJ074836.75+521319.9& 24.0      &  11354    \\
1996bl     &  6.7& WISEAJ003617.70+112342.6& 24.8      &  19235    \\
%	2012cf &10.3    &CGCG 161-119&49.9&RC3\\
	2000fs&0.0&NGC 1218&84.8&RC3\\
      2014ap    &  8.6&      CGCG 126-07& 39.0& 15138\\
      2013fj    &  5.8&      CGCG 428-06& 21.8& 1702\\
      2013ey    &  3.6&      CGCG 425-02& 26.8& 4441\\
      2012eo    &  9.2&      CGCG 494-00& 32.8& 8558\\
      2012dg    & 10.0&      CGCG 254-03& 30.4& 12531\\
      2010ew    &  8.6&      CGCG 173-01& 20.2& 4740\\
      2006te    &  6.3&      CGCG 207-04& 27.0& 3927\\
      2005bu    &  8.1&      CGCG 117-01& 29.2& 10362\\
      2004ck    & 10.6&      CGCG 064-09& 20.2& 6426\\
      2002bf    &  4.1&      CGCG 266-03& 38.6& 12138\\
      2002aw    &  2.8&      CGCG 189-02& 32.0& 13736\\
      2000fo    &  8.5&      CGCG 475-01& 39.4& 12654\\
      2000df    &  9.5&      CGCG 051-07& 24.4& 2381\\
      2000cv    & 10.0&      CGCG 292-08& 22.8& 3297\\
      1996ac    &  3.6&      CGCG 014-02& 21.8& 6452\\
      SN2014K   &  7.1&      CGCG 332-20& 22.4& 3457\\
      SN2012B   &  9.4&      CGCG 526-10& 24.2& 5199\\
      SN2010V   &  7.8&      CGCG 163-59& 34.2& 3819\\
      SN1999X   &  7.2&      CGCG 180-22& 28.8& 7507\\
      SN1994Q   &  4.0&      CGCG 224-10& 21.2& 10504\\

\hline
 %2000cv      RC3 17.7 /     K_s 22.8 /     K_s 43.1 /    RC3 17.7 /   
\end{tabular}
\end{table}
%\end{document} 

\section*{References}
\noindent Barnes, R. 2001,  "Gravitational Focusing", in Gargaud, Amils, Quintanilla, Cernicharo, \& Henderson eds., Encyclopedia of Astrobiology, Berlin, Heidelberg: Springer, p. 692\\ 
Bell E. F., McIntosh D. H., Katz N., Weinberg M. D., 2003, ApJS, 149, 289 \\
Blanton M. R. \etal 2003, ApJ, 592, 819\\
Bernyk, M. \etal 2016, ApJS, 223,9\\
Binney, J. \& Vasiliev, E. 2023, MNRAS, 520, 1832\\
Cole, D. \& Binney, J. 2016, MNRAS, 465, 798\\
Croton, D. \etal 2023, http://dx.doi.org/10.26185/642bbd010a9ca\\
de Vaucouleurs, G. 1959, Handbuch der Physik (Springer-Verlag,
Berlin, G\"ottingen), Vol. 53.\\
de Vaucouleurs, G., de Vaucouleurs, A. \& Corwin 1976, Reference Catalog of Bright Galaxies\\
Desai, D., Shappee, B. Kochanek, C. \etal 2026,  arXiv 2602.00223\\
Freeman, K. 1970, ApJ, 160, 811\\
Graham, P., Rajendran, S. \& Varela, J. 2015, PhRvD, 92f3007\\%Dark matter triggers of supernovae
Lazar, A. \etal 2020, MNRAS, 497, 2393\\
Leung, Shing-Chi, Walther, S., Nomoto, K. \& Kusenko, A. 2025, ApJ, 991, 11\\
Riess, A. \etal 2024, ApJ, 977, 120\\
Seligman, D., Marceta, D. \& Pe\~na-Asensio, E. 2026, ApJ, 997, 146\\
Springel V. \etal 2005, Nature, 435, 629\\
Toy, M. \etal 2023, MNRAS, 526, 5292\\
Troxel, M. \etal 2018, PhRvD, 98d3528\\
%Wegg, C. \& Gerhard, O. 2013, MNRAS, 435, 1874\\
%Quezada, C. \etal 2025, A\&A, 702, 164\\
%Mould, J. 1983, ApJ, 266, 255\\
%Freeman, K., de Vaucouleurs, G., de Vaucouleurs, A. \& Wainscoat, R. 1988, ApJ, 325, 563\\
%Maguire, K., Eracleous, M.,  Jonker, P., MacLeod, M. \& Rosswog, S. 2020, SSRv, 216, 39\\
\section*{Appendix}%\appendix
\renewcommand{\thetable}{A\arabic{table}}
\renewcommand{\thefigure}{A\arabic{figure}}
\setcounter{figure}{0}
\setcounter{table}{0}
To compare 2MASS isophotal diameters with
RC3 D$_{25}$ diameters we show the ratio in Figure A1 for those galaxies that have both. The mean value and the median are 0.690 $\pm$ 0.008.
\begin{figure}
    \includegraphics[width=.75\textwidth]{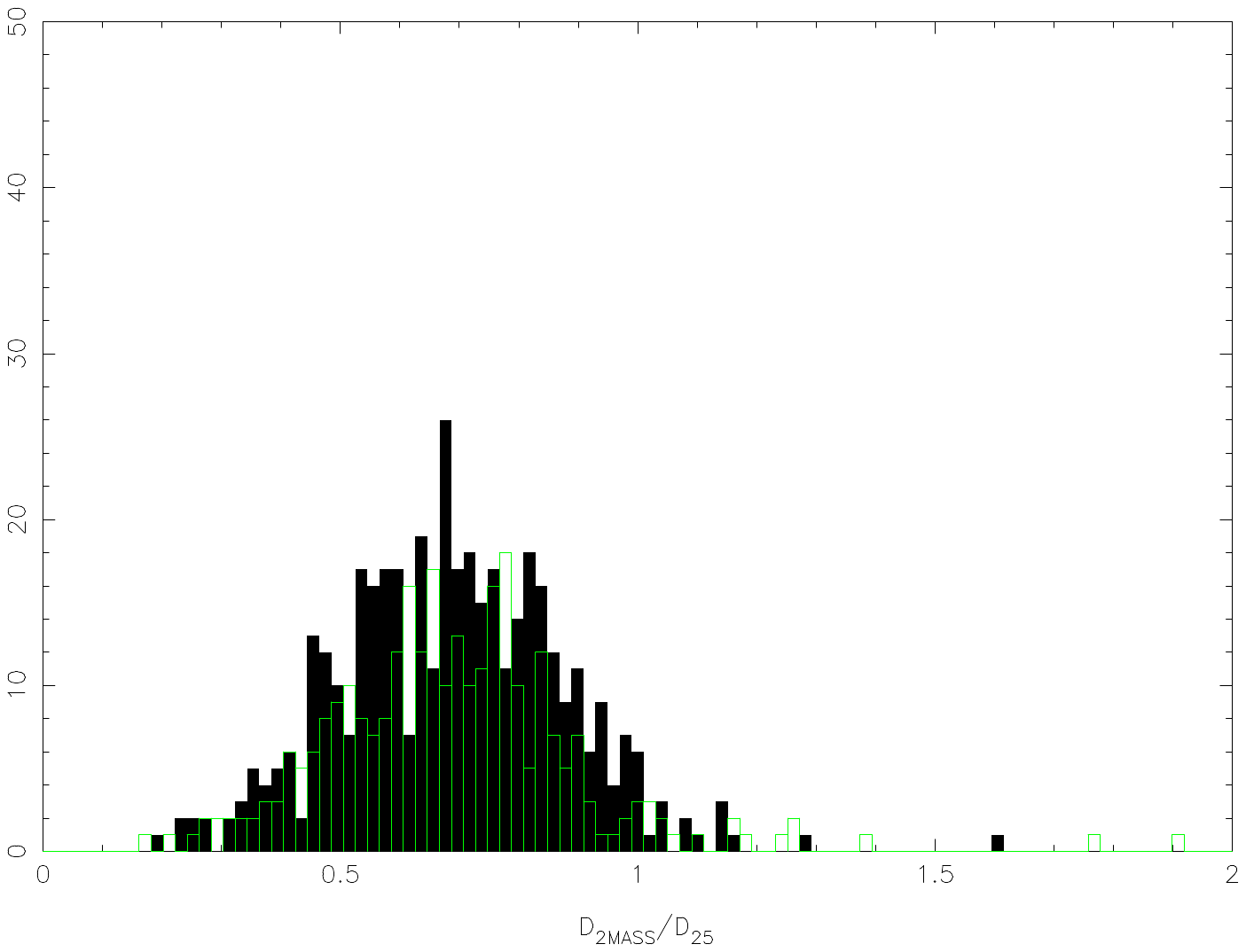}
    \caption{Ratio of 2MASS isophotal diameters to RC3 D$_{25}$ diameters. There is no noticeable difference between the black z $<$ 0.02 subsample and the green z $>$ 0.02 one. }
%    \end{figure}
\end{figure}

Surface brightness dimming affects galaxy diameter measurements.
In magnitudes the effect is 2.5$\log$ (1+z)$^4$ = 10$\log$(1+z), where z is the redshift.
A value of D$_{25}$ can be corrected by multiplying by 3.4/(3.4-10z) for z$<<$1,
3.4 being the difference between the 25th mag isophote and the central surface brightness of a Freeman disk: 21.6 mag arcsec$^{-2}$.
%The median ratio of RC3 diameter to 2MASS total magnitude diameter is 1.4.
In Figure A2 we show the effect on the highest redshift sample, which is Table 2.
The correction factor was %A value of D$_{25}$ can be corrected by multiplying by 
2.346/(2.346-10z) for z$<<$1.
\begin{figure}
	\includegraphics[width=.75\textwidth]{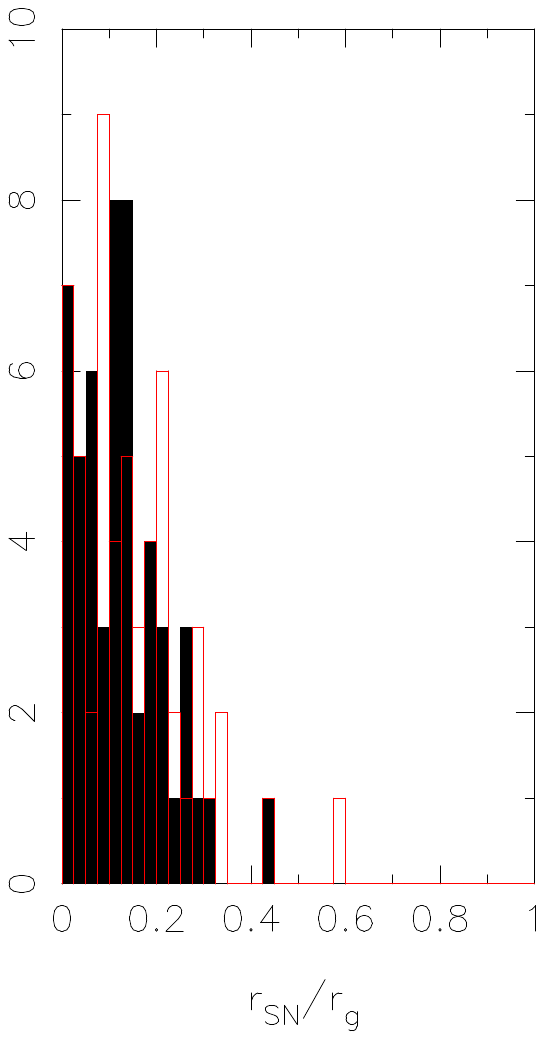}
	\caption{Raw SN radii divided by 2MASS radius (red histogram).
	The solid black histogram has been corrected for surface brightness dimming.}
\end{figure}

%\keywords{\uat{Galaxies}{573} --- \uat{Cosmology}{343} --- \uat{High Energy astrophysics}{739} --- \uat{Interstellar medium}{847} --- \uat{Stellar astronomy}{1583} --- \uat{Solar physics}{1476}}


\begin{acknowledgments}
We thank all the people that have made this AASTeX what it is today.  This
includes but not limited to Bob Hanisch, Chris Biemesderfer, Lee Brotzman,
Pierre Landau, Arthur Ogawa, Maxim Markevitch, Alexey Vikhlinin and Amy
Hendrickson. Also special thanks to David Hogg and Daniel Foreman-Mackey
for the new {\tt\string modern} style design. Considerable help was provided via bug
reports and hacks from numerous people including Patricio Cubillos, Alex
Drlica-Wagner, Sean Lake, Michele Bannister, Peter Williams, Jonathan
Gagne, Arthur Adams, Nicholas Wogan, Aaron Pearlman, Jeff Mangum, Mark Durre, Joel Ong, and Stephen Thorp.
\end{acknowledgments}

\begin{contribution}
%%This section gives authors the space to recognize author contributions. The text inside this environment is NOT counted towards the total word quanta. At a minimum, manuscripts are expected to include this text:

All authors contributed equally to the Terra Mater collaboration.

%% But authors are expected to provide more specific details, e.g. 
%%
%%SC was responsible for writing and submitting the manuscript.
%%WWM came up with the initial research concept and edited the manuscript.
%%OTS obtained the funding and edited the manuscript.
%%EBF provided the formal analysis and validation. He also edited the manuscript.
%%GEH Supervised the undergraduates, wrote the software and administers the project github and Zenodo repositories.
%%
%% Authors can use the Contributor Role Taxonomy (CRediT) at
%% for ideas on how write a good statement tailored to their needs.

\end{contribution}

\facilities{HST(STIS), Swift(XRT and UVOT), AAVSO, CTIO:1.3m, CTIO:1.5m, CXO}

\software{astropy \citep{2013A&A...558A..33A,2018AJ....156..123A,2022ApJ...935..167A},  
          Cloudy \citep{2013RMxAA..49..137F}, 
          Source Extractor \citep{1996A&AS..117..393B}
          }


\appendix

\section{Appendix information}

Appendices can be broken into separate sections just like in the main text.
The only difference is that each appendix section is indexed by a letter
(A, B, C, etc.) instead of a number.  Likewise numbered equations have
the section letter appended.  Here is an equation as an example.
\begin{equation}
I = \frac{1}{1 + d_{1}^{P (1 + d_{2} )}}
\end{equation}
Appendix tables and figures should not be numbered like equations. Instead
they should continue the sequence from the main article body.

\section{Author publication charges} \label{sec:pubcharge}

In April 2011 the traditional way of calculating author charges based on 
the number of printed pages was changed.  The reason for the change
was due to a recognition of the growing number of article items that could not 
be represented in print. Now author charges are determined by a number of
digital ``quanta''.  A single quantum is defined as 350 words, one figure, one table,
and one digital asset.  For the latter this includes machine readable
tables, data behind a figure, figure sets, animations, and interactive figures.  The current cost
for the different quanta types is available at 
\url{https://journals.aas.org/article-charges-and-copyright/#author_publication_charges}. 
Authors may use the ApJL length calculator to get a {\tt rough} estimate of 
the number of word and float quanta in their manuscript. The calculator 
is located at \url{https://authortools.aas.org/ApJL/betacountwords.html}.

\section{Rotating tables} \label{sec:rotate}

To place a single page table in a landscape mode start the table portion with {\tt\string\begin\{rotatetable\}} and end with {\tt\string\end\{rotatetable\}}.

Tables that exceed a print page take a slightly different environment since both rotation and long table printing are required. In these cases start with {\tt\string\begin\{longrotatetable\}} and end with {\tt\string\end\{longrotatetable\}}. The {\tt\string\movetabledown} command can be used to help center extremely wide, landscape tables. The command {\tt\string\movetabledown=1in} will move any rotated table down 1 inch. 

A handy "cheat sheet" that provides the necessary \latex\ to produce 17
different types of tables is available at \url{http://journals.aas.org/authors/aastex/aasguide.html#table_cheat_sheet}.

\section{Using Chinese, Japanese, and Korean characters}

Authors have the option to include names in Chinese, Japanese, or Korean (CJK)
characters in addition to the English name. The names will be displayed
in parentheses after the English name. The way to do this in AASTeX is to
use the CJK package available at \url{https://ctan.org/pkg/cjk?lang=en}.
Further details on how to implement this and solutions for common problems,
please go to \url{https://journals.aas.org/nonroman/}.

%% For this sample we use BibTeX plus aasjournalv7.bst to generate the

\bibliography{sample701}{}
\bibliographystyle{aasjournalv7}



\end{document}